\documentclass[10pt]{article}
\usepackage[utf8]{inputenc}
\usepackage[T1]{fontenc}
\usepackage[english]{babel}
\usepackage{csquotes}
\usepackage{amsmath,amsfonts,amssymb}
\usepackage{graphicx,xcolor}
\usepackage{booktabs}
\usepackage{microtype}
\usepackage{enumitem}
\usepackage[most]{tcolorbox}

\usepackage[a4paper, left=2cm, right=2cm, top=2.25cm, bottom=2.25cm]{geometry}

\definecolor{light_blue}{RGB}{179, 207, 251}
\definecolor{mint}{RGB}{115, 255, 187}

\usepackage{lmodern}
\usepackage{times}
\usepackage{mathptmx}

\usepackage[labelfont={small,bf,sf}, labelsep=period, format=plain, justification=justified]{caption}

\usepackage{titlesec}
\titleformat{\section}
  {\sffamily\Large\bfseries}  
  {\thesection}{1em}{}
\titleformat{\subsection}
  {\sffamily\large\bfseries}  
  {\thesubsection}{1em}{}
\titleformat{\subsubsection}
  {\sffamily\normalsize\bfseries}  
  {\thesubsubsection}{1em}{}
\titleformat{\paragraph}[runin]{\sffamily\small\bfseries}{}{0pt}{}[.]

\titlespacing*{\section}{0pt}{24pt}{4pt}
\titlespacing*{\subsection}{0pt}{16pt}{2pt}
\titlespacing*{\subsubsection}{0pt}{12pt}{2pt}
\titlespacing*{\paragraph}{0pt}{8pt}{10pt}

\newcommand{\sectionheader}[1]{
  \phantomsection
  \section*{#1}  
  \addcontentsline{toc}{section}{#1}  
}
\newcommand{\subsectionheader}[1]{
  \phantomsection
  \subsection*{#1}  
  \addcontentsline{toc}{subsection}{#1}  
}

\usepackage{authblk}
\makeatletter
\renewcommand{\maketitle}{
  \begin{flushleft}    
    \sffamily    
    {\fontsize{25}{32}\fontfamily{phv}\selectfont\bfseries\@title\par}
    \vskip 1.6em
    {\fontsize{12}{16}\selectfont\@author\par}    
  \end{flushleft}
}
\makeatother

\makeatletter
\renewenvironment{abstract}{     
    \small
    \sffamily
    \vskip 1.0em
    \noindent\rule{\textwidth}{0.2pt}
    \vskip 1.0em
    \begin{quote}
        \fontfamily{lmss}\selectfont
        \setlength{\leftskip}{-1.9em}
        \setlength{\rightskip}{-1.0em}
        \large\bfseries\abstractname\vspace{-.5em}\par\bigskip        
        \normalfont\sffamily\small
}{
    \end{quote}    
    \noindent\rule{\textwidth}{0.2pt}
    \vspace{-2pt}
}
\makeatother

\usepackage[backend=biber, hyperref=true, natbib=true, doi=true, url=false, eprint=true, date=year, clearlang=true, style=nature, sorting=none, maxnames=100, minnames=100]{biblatex}
\DeclareFieldFormat{labelnumberwidth}{\textit{#1.}}

\usepackage[breaklinks]{hyperref} 
\hypersetup{colorlinks=true, citecolor=blue, linkcolor=blue, urlcolor=blue}

\usepackage{fancyhdr}  
\usepackage{lastpage}
\title{Risks of AI-driven product development and strategies for their mitigation}

\author[1,2,*]{Jan Göpfert}
\author[1]{Jann M. Weinand}
\author[1]{Patrick Kuckertz}
\author[1]{Noah Pflugradt}
\author[1]{Jochen Linßen}
\affil[1]{Institute of Climate and Energy Systems, Jülich Systems Analysis (ICE-2), Forschungszentrum Jülich, 52425 Jülich, Germany}
\affil[2]{Chair for Fuel Cells, Faculty of Mechanical Engineering, RWTH Aachen University, 52062 Aachen, Germany}

\affil[*]{corresponding author: Jan Göpfert (j.goepfert@fz-juelich.de)}

\begin{document}

\flushbottom
\maketitle

\thispagestyle{empty}

\begin{abstract}
Humanity is progressing towards automated product development, a trend that promises faster creation of better products and thus the acceleration of technological progress. However, increasing reliance on non-human agents for this process introduces many risks. This perspective aims to initiate a discussion on these risks and appropriate mitigation strategies. To this end, we outline a set of principles for safer AI-driven product development which emphasize human oversight, accountability, and explainable design, among others. The risk assessment covers both technical risks which affect product quality and safety, and sociotechnical risks which affect society. While AI-driven product development is still in its early stages, this discussion will help balance its opportunities and risks without delaying essential progress in understanding, norm-setting, and regulation.
\end{abstract}

\sectionheader{Introduction}
Imagine AI agents increasingly automating the product development process, eventually leading us to a future in which access to large engineering capabilities lies at our fingertips. The demand for human ingenuity and expertise in the development of complex products would decrease, whilst technological progress would be largely driven by algorithms. While this is still science fiction, recent advances in AI are pushing us strongly in this direction, redirecting large sums of research and industry money towards such visions, as evidenced by the rapidly growing number of publications at the intersection of AI and engineering (see Figure~\ref{fig:ai_and_engineering_literature_growth}). 
This future scenario is also acknowledged by the ongoing discussion within the World Intellectual Property Organisation (WIPO) about the impact of AI-generated inventions on the patent system~\supercite{wiposecretariatRevisedIssuesPaper2020}. Albeit we are very much at the beginning of this process, we have no doubt that humanity is moving towards higher levels of automation in product development and that it is worth starting to discuss the implications of that. \textit{What are the social, economic, and ethical risks? What are regulatory and legal challenges? What technical threats arise and how can we `trust' products and technologies that are placed in our hands from an opaque black box?}

Large language models (LLMs) are powerful text generation `black boxes' that fuel the current stark progress in AI. Although we know their math and gain much understanding from empirical evidence, they are too large and complex for their inner workings to be made fully transparent and comprehensible~\supercite{jainAttentionNotExplanation2019,mosbachInsightsActionsImpact2024,anwarFoundationalChallengesAssuring2024,sharkeyOpenProblemsMechanistic2025}. Today, the principles behind LLMs extend to other modalities---from text, to vision, to audio~\supercite{nanFrontierReviewMultimodal2023}. Furthermore, adding a layer of complexity on top of models we already do not fully understand, the models become integrated in frameworks in which many consecutive model calls allow for agentic behavior such as planning, acting, memorization, and self-reflection~\supercite{parkGenerativeAgentsInteractive2023}. 
This technology is advancing rapidly, as exemplified by the exponential increase in the length of software tasks that LLM-based agents can perform with a certain level of reliability~\supercite{kwaMeasuringAIAbility2025}.
By definition, the advent of general-purpose technologies has had broad impacts on our societies~\supercite{bresnahanGeneralPurposeTechnologies1995}, including on the way products are designed and manufactured. Historically, the printing press changed the sharing of engineering knowledge; the steam engine brought mass production; the computer led to simulation-driven and computer-aided design~(CAD); and the internet gave rise to cloud computing, digital twins, and transformed the way we collaborate. Arguably, large language models and other foundation models are such a general-purpose technology~\supercite{wiposecretariatRevisedIssuesPaper2020,craftsArtificialIntelligenceGeneralpurpose2021,bommasaniOpportunitiesRisksFoundation2022a}, and therefore introduce transformative changes to the product development process.

\begin{figure}
    \centering
    \includegraphics[width=0.96\linewidth]{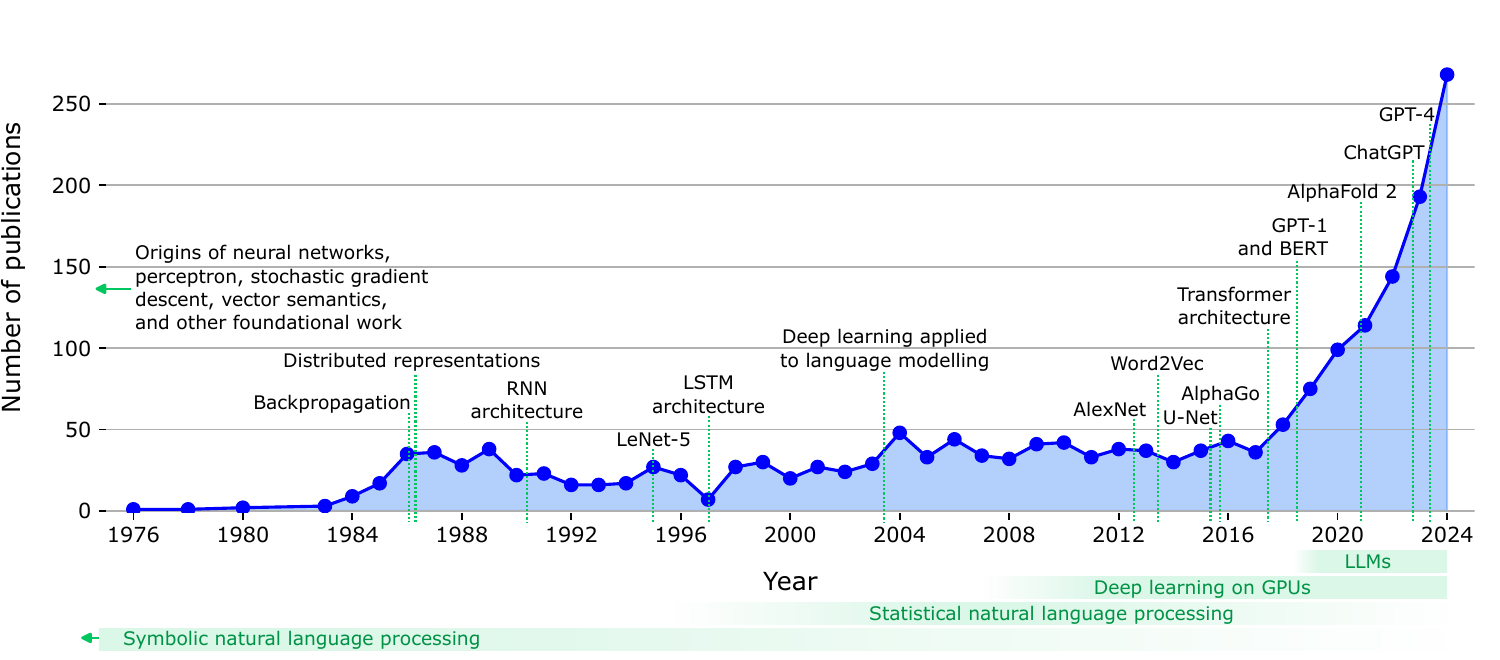}
    \caption{Annual documents published on the intersection of AI and engineering with historical milestones and paradigms in AI highlighted. The annual number of publications began to rise sharply around 2017, increasing sevenfold by 2024. See Jurafsky and Martin~\supercite{jurafskySpeechLanguageProcessing2025} and Gururaja et al.~\supercite{gururajaBuildOurFuture2023} for historical overviews on AI from the perspective of natural language processing. The data is from Scopus using the following query:      
    \texttt{\footnotesize TITLE-ABS-KEY(("AI"  OR  "artificial intelligence") AND ("product development" OR "engineering design")) AND (LIMIT-TO(SUBJAREA, "ENGI") OR LIMIT-TO(SUBJAREA, "COMP"))}.}
    \label{fig:ai_and_engineering_literature_growth}
\end{figure}

Product development is a broad term covering many industries. Today's `cyber-physical' products often combine knowledge from mechanical, electrical, and software engineering as well as other disciplines. In this paper, we mainly consider the development of technical products that have a spatial dimension and may or may not contain software (i.e., machines and devices). Pure software development is not in scope, but learnings from its already high adoption of AI assistance are valuable. Most software developers are now using AI tools~\supercite{stackoverflowAI2024Stack, klemmerUsingAIAssistants2024} and see increased productivity as the greatest benefit~\supercite{stackoverflowAI2024Stack, klemmerUsingAIAssistants2024}, while several studies suggest a reduction in secure code~\supercite{pearceAsleepKeyboardAssessing2022,perryUsersWriteMore2023,majdinasabAssessingSecurityGitHub2024}. 

This paper provides a perspective on the risks associated with the increasing automatization of the product development process using opaque black boxes and corresponding mitigation strategies. The risks heavily depend on the design and usage of the systems; are they designed to alert human engineers to potential errors, or are they designed to develop products autonomously? In the former case, the number of design errors would presumably decrease and the risks are low. In the latter case, the answer depends on the quality of the system, and the risks are high, including those whose solvability is questionable, such as insufficient alignment~\supercite{wolfFundamentalLimitationsAlignment2025}.
In this perspective, we consider the more extreme and distant case of AI systems that perform the design task and optionally other tasks with high autonomy to derive associated risks and mitigation strategies.
Among the many risks associated with the development and use of AI, there are arguments  for the presence of existential risks (e.g., from self-improving and power-seeking AI)~\supercite{balesArtificialIntelligenceArguments2024,scienceinnovationandtechnologycommitteeGovernanceArtificialIntelligence2023}. This is beyond the scope of this paper. Instead, we focus on risks that are more tangible from today's perspective.

The limitations of LLMs and corresponding vulnerabilities of applications relying on them as well as mitigation strategies have been widely studied~\supercite{bommasaniOpportunitiesRisksFoundation2022a,weidingerTaxonomyRisksPosed2022,kirkBenefitsRisksBounds2024,klemmerUsingAIAssistants2024,anwarFoundationalChallengesAssuring2024,gyevnarAISafetyEveryone2025}. Individual research domains have begun consensus-building processes to assess and manage the risks of AI usage in their fields~\supercite{tzachorResponsibleArtificialIntelligence2022,trotsyukFrameworkRiskMitigation2024}. The path towards automated product development is highlighted by recent publications on the application of generative AI to mechanical system configuration design~\supercite{etesamDeepGenerativeModel2024}, requirements elicitation~\supercite{ataeiElicitronLLMAgentBased2024}, optimization~\supercite{picardGenerativeOptimizationPerspective2024}, and extracting and complying to requirements in technical documentation~\supercite{dorisDesignQAMultimodalBenchmark2024}. AI is expected to complement and integrate existing computational methods in product development (e.g., simulations) by participating in the creative reasoning process previously reserved for humans~\supercite{gopfertOpportunitiesLargeLanguage2024}. Literature in technology and innovation management focuses on the value and effective implementation of AI in organizational contexts, e.g., by providing guidance on trust~\supercite{pillerGenerativeAIInnovation2024} and effective human-machine collaboration~\supercite{pillerHybridIntelligenceInnovation2023,bouscheryArtificialIntelligenceAugmentedBrainstorming2024}.
Product safety is safeguarded by national and international regulations (e.g., EU's `GPSR'~\supercite{euRegulationEU20232023}) and standards, some of which are sector-specific (e.g., for medical devices or aircraft). Compliance must be ensured by economic operators (e.g., manufacturers, who are therefore required to have internal processes for product safety) and is monitored by market surveillance authorities. Compliance with national and international standards, certification schemes, and sector-specific codes of good practice can, among other things, assist in risk assessment~\supercite{euRegulationEU20232023}.
Safety engineering is an engineering discipline concerned with the safety of technical systems.
The study of potential implications of new and emerging technologies is known as technology assessment~\supercite{grunwaldTechnologyAssessmentConcepts2009}, and concerns about AI in particular are discussed in the AI safety community~\supercite{gyevnarAISafetyEveryone2025}.
The regulation of emerging technologies is studied across multiple disciplines (i.a., law, political science, and economics). The risks posed by new technologies were modeled in terms of their adoption, finding that socially optimal adoption is initially slow and gradually increasing, with slower adoption rates reducing the risk of catastrophic outcomes~\supercite{acemogluRegulatingTransformativeTechnologies2024}. 
To limit negative impacts of AI, institutional, national and global norms are being negotiated, and a plethora of guidelines and principles have been published by government agencies, supranational organizations, academic institutions, non-profit organizations, and private companies alike~\supercite{jobinGlobalLandscapeAI2019}. Governance of AI, while still in its infancy, is attracting increasing research and policy attention~\supercite{tallbergGlobalGovernanceArtificial2023}. The European Union's AI Act~\supercite{euRegulationEU20242024}, proposed in 2021 and enacted in 2024, is regarded as the first far-reaching regulation of AI in a major jurisdiction~\supercite{schuettRiskManagementArtificial2024}. 

While the limitations of AI in product development have been briefly touched upon in some cases, they have never been a focus. To our knowledge, this is the first assessment of the risks and corresponding mitigation strategies associated with the increasing use of AI in product development. By raising awareness and proposing basic principles for safe AI-driven engineering, we provide an important perspective on this emerging transformation, as well as a unique perspective on AI safety and ethics.

\sectionheader{Risks}
In this section, we provide an overview of 1) concerns about the quality and correctness of designs generated by automated product development systems, and 2) societal, ethical, environmental and legal concerns. It should be noted that these categorizations are not mutually exclusive. A technical risk often entails a social risk on an aggregated level. 
Although we refer to systems tailored for product development, the risks also apply to general-purpose AI models that may be used in similar ways. For brevity, we mainly focus on the risks associated with the design task. However, it is likely that AI tools will be used throughout the whole product development process. Therefore, the risks are not only present in solving engineering problems, but also in upstream and downstream tasks such as requirements definition.
For example, software developers are already using AI tools for requirements analysis and documentation~\supercite{klemmerUsingAIAssistants2024}, and in product design the use of LLMs for requirements elicitation is being explored~\supercite{ataeiElicitronLLMAgentBased2024}. 
Not all development activities are equally exposed to the risks described below. The severity of the risks depends on the product or technology being developed~(e.g., pen vs. plain), its application~(e.g., medical device vs. decorative item), the design stage and its proximity to production~(e.g., preliminary design vs. detailed design), the AI system~(e.g., co-pilot vs. fully autonomous agent), and how it is used~(e.g., with or without human validation).

\subsectionheader{Technical concerns} 
Systems for automatic product development will output engineering solutions that contain \textbf{errors}---presumably plenty at the beginning and fewer with further progression of these systems, assuming identical usage. With decreasing number of errors, it is likely that also human oversight decreases, lowering the probability of an individual error to be spotted. In addition, human decision-making is subject to automation bias~\supercite{skitkaDoesAutomationBias1999}. 
The ability of today's LLMs to convincingly make false statements~\supercite{spitaleAIModelGPT32023} suggests that engineers will be prone to accepting erroneous results from AI systems. For example, studies indicate that software developers are overconfident about the security of their code when using AI tools~\supercite{perryUsersWriteMore2023, klemmerUsingAIAssistants2024}. Even if the generated solutions do not contain errors, insufficient system capabilities will lead to \textbf{suboptimal design} decisions (e.g., rejecting a favorable solution), which is exacerbated by the biases inherent in the systems.

Besides bad execution of a design task, another problem resulting in flawed designs is \textbf{misalignment} between the design objectives as interpreted by the AI system and the human intentions. Alignment is a known and difficult to solve issue~\supercite{wolfFundamentalLimitationsAlignment2025}. In particular, latent requirements that seem obvious to us and are difficult to express, may not be considered. 
Arguably, design that is intended to benefit people must take human values~\supercite{friedmanValuesensitiveDesign1996} and judgment into account.
While using AI to capture user needs could strengthen the user perspective in product development~\supercite{mariontuckerj.AIUserCenteredNew2023}, applying it towards automating design could lead to a \textbf{disconnect} between product development and the users the product should benefit, resulting in poor design decisions.
Furthermore, the operation of such systems in a too \textbf{narrow context} may lead to inadequate or harmful products and technologies. 
Technologies are situated in a complex, sensitively balanced world, affecting societies, economies, ecosystems, and the environment. If AI is used to develop new products and technologies, these complex surroundings and their interdependencies must be considered. 

Another reason for AI not executing the design task as intended is \textbf{targeted manipulation}. AI systems such as LLMs are vulnerable to adversarial attacks, training data poisoning, and more~\supercite{zouUniversalTransferableAdversarial2023,qiSafetyAlignmentShould2024}. For example, prompt injections---purposefully or accidentally---can lead to unexpected behavior (e.g., the creation of a backdoor or a bias towards a particular supplier). Similarly, malicious users could circumvent measures for safety and ethical compliance (referred to as jailbreaking) making it difficult to prevent dual-use of automatic product development systems.
In addition to a malicious user or third party deliberately exploiting the system to cause harm, the system itself can be the source of harm (e.g., by leaking confidential information or concealing failure modes). The increasing trend towards agent-based AI systems carries the risk of them autonomously performing harmful actions---triggered by a malicious actor or accidentally (e.g., as a result of reward hacking~\supercite{skalseDefiningCharacterizingReward2022}). \textbf{Excessive agency} can occur if such systems are not sufficiently constrained in functionality, permissions or autonomy~\supercite{owaspllmprojectLLM062025Excessive}. 
The ability to autonomously execute potentially long chains of actions makes ensuring the safety of AI agents challenging, especially in multi-agent environments~\supercite{chanHarmsIncreasinglyAgentic2023,anwarFoundationalChallengesAssuring2024}.

\subsectionheader{Sociotechnical concerns}
As described above, technologies cannot be assessed in isolation, as they form a dynamic system with the broader context in which they are embedded. Consequently, their risks are not only technical, but include social, ethical, environmental, and legal dimensions, among others.
By potentially accelerating innovation whilst re-distributing inventive power and labor, the introduction of systems that automate product development would have large effects on our socioeconomic system. 
Uncertainty in predicting possible implications is a risk in itself. 
The unknowns include whether certain positive effects will outweigh the opposing negative effects or vice versa and what \textbf{indirect risks} will arise. For example, adding to the environmental cost of the energy-intensive training and use of AI models~\supercite{strubellEnergyPolicyConsiderations2019,pattersonCarbonEmissionsLarge2021,vriesGrowingEnergyFootprint2023}, the potential acceleration of innovation may result in shorter product iterations which may lead to increased consumption and consequent environmental damage. Conversely, the systems may lead to more environmentally compatible innovations resulting in an overall positive effect.

What is certain is that systems for automated product development would constitute a powerful technology that is subject to the `\textbf{dual-use dilemma}' enabling both legitimate and reprehensible use cases. It is safe to assume that those systems will only be adopted if they increase engineering capabilities and/or lower development costs. In either case, the risk of harmful technology being created is raised by increasing the amount of technical problems that can be solved and/or the group of players having access to a certain level of engineering capabilities, assuming regulations remain unchanged. The dual-use dilemma of AI is widely recognized  with regard to the development of chemical, biological, radiological, and nuclear weapons~\supercite{nistManagingMisuseRisk2024,trotsyukFrameworkRiskMitigation2024,anwarFoundationalChallengesAssuring2024}. Even though the classical disciplines of engineering such as mechanics, fluid dynamics, thermodynamics, heat and mass transfer, or electrical engineering are not congruent with chemical, biological, radiological, and nuclear knowledge, they nevertheless facilitate its application, not least by allowing the development of necessary technical peripherals. 

If product development were to become inexpensive, it might enable the creation of unethical products that would not have been feasible in traditional, cost-intensive, corporate-driven, and image-dependent development. 
Furthermore, at the national level, it is arguably easier to switch from developing civilian to military applications when there is less need for human expertise, whose specializations are slow to change. Lastly, it can be argued that companies will be \textbf{less sensitive to unethical behavior} if they replace human engineers, who have a moral compass and can criticize or expose unethical behavior, with computer systems. Also, by reducing the size of development teams, their diversity may decrease. Though biases are already present in current development departments, the \textbf{biases} introduced by AI systems are potentially more dangerous as they are global and systematic.

The \textbf{threats to accountability} posed by computers and algorithms equally apply to `computerized engineering'. These are the diffusion of responsibility through the problem of many hands, the thinking of bugs as inevitable, blaming computers instead of legal entities, and software ownership without liability~\supercite{nissenbaumAccountabilityComputerizedSociety1996,cooperAccountabilityAlgorithmicSociety2022}. Arguably, this threat is increasing---from calculators to engineering AI agents---as automation systems become more complex and less deterministic in nature, thereby reducing traceability of errors to user input.

Skill variety, skill use, task identity, task significance, autonomy, social contact, and feedback are important factors in meeting \textbf{workers' needs} and work outcomes~\supercite{hackmanMotivationDesignWork1976,parkerOneHundredYears2017}, but are challenged by job automation~\supercite{parkerAutomationAlgorithmsWhy2022}.
For example, shifting from creative to judgmental activities and rendering parts of professional training superfluous can reduce job satisfaction~\supercite{toner-rodgersArtificialIntelligenceScientific2024}. The presence of AI and associated digital work environments, along with more structured, comparable task design, may lead to more comprehensive monitoring of employee performance~\supercite{mettlerConnectedWorkplaceCharacteristics2024,acemogluHarmsAI2021}.
Furthermore, a \textbf{displacement of labor tasks}~\supercite{Cazzaniga2024Gen} can lead to significant social and political tensions, lowering well-being~\supercite{nazarenoImpactAutomationArtificial2021,giuntellaArtificialIntelligenceWorkers2023}, and reduce workers' bargaining power~\supercite{estlundLosingLeverageEmployee2023}.

The risk of \textbf{concentration of power} goes beyond its shift from workers to organizations: Access to the means of generating innovation is not equally distributed, and their automation may exacerbate these inequalities~\supercite{Cazzaniga2024Gen}. The potential acceleration of innovation by these systems may further reinforce these inequalities. The inequality relates to both access to and provision of such systems and is reflected in global disparities (e.g., between the global North and South) and market concentration. Conversely, if access to said systems becomes inexpensive, it could have the opposite effect and balance out technical development capabilities around the world. 
Excessive market power among providers of such systems would entail high risks, as monopolizing technological progress is qualitatively different from monopolizing a specific technology such as smartphones. Digital services have strong centralizing tendencies, as they are globally available and easily scalable, and sometimes have self-reinforcing properties such as the `network effect'~\supercite{fletcherEffectiveUseEconomics2024}; broad access to development data and user interaction could self-reinforce a market leader's position.
Even with many competing providers of AI systems for engineering, they may be dependent on a few players who can afford to competitively train and host the underlying models and thereby withhold fundamental ``editorial decisions,'' analogous to the current AI landscape~\supercite{widderWhyOpenAI2024}.

Without adequate security measures, an increasing digitization carries the risk of more sensitive \textbf{data breaches}. When asked about security, participants in a survey on AI-assisted software development were most concerned about privacy, not the quality of their work~\supercite{klemmerUsingAIAssistants2024}. Systems for automatic product development will likely have access to sensitive business data which may be disclosed to not authorized business units or third parties. While this is true of today's software, it likely has less comprehensive access and lacks agency. In the case of centralized systems, their providers have enormous power and responsibility, holding product development data from many companies across the globe. The providers of such services may be forced to give their government access to their data or third parties unlawfully gain access. 
\textbf{Dependence} of companies and nations on a few system providers could make economies fragile and vulnerable. In the context of hybrid warfare, the automation of development capabilities makes them attractive targets for nation-state hacking, especially in the case of centralized systems.

Another issue is the accompanying change in \textbf{engineering education and research}, which are already dependent on access to expensive resources (e.g., test benches and simulation software). The advent of automated product development systems could increase these dependencies and widen the gap between institutions of different wealth and between the academic and industrial sector.
Whereas previously the cost of the technical artifact being studied could create an imbalance (e.g., the cost of developing cars is a higher barrier to research for automotive engineering institutes than for car manufacturers), now the ownership of AI systems by corporate actors could create barriers to studying and improving the product development process itself and the tools supporting it. 
A parallel can be drawn with natural language processing research, where LLMs were pioneered and studied, but their scaling put the resources necessary to train these models beyond the reach of most academic organizations, making them dependent on models from mostly corporate actors, thereby leading to concerns about the ``secrecy of industry research'', the scientific value of research based on non-transparent models, having to compete with Big Tech for research to be valued, and ``building an upper class of AI''~\supercite{gururajaBuildOurFuture2023}.

There are also legal challenges. For example, the \textbf{patent system} rewards the disclosure of one's own inventions of a certain threshold of inventiveness, novelty, and commercial applicability through intellectual property rights. 
While the patent system is already criticized for granting patents to rather obvious ``inventions'' and for other issues~\supercite{boldrinCasePatents2013}, if automated product development systems become a reality, they would exacerbate this problem by lowering the barrier to innovation, thus requiring lower rewards or higher levels of inventiveness.
In an extreme case, the patent system is led ad absurdum. While automatic product development systems may undermine the patent system, in view of the dual-use dilemma, the `registration of inventions' may become more important from a security perspective in order to control the ``democratized'' inventive power. The implications of AI-generated inventions for the patent system and how it should be adapted are the subject of ongoing debate~\supercite{wiposecretariatRevisedIssuesPaper2020,ramalhoPatentabilityAIGeneratedInventions2018,derassenfosseAIGeneratedInventionsImplications2023}.

\sectionheader{Mitigation strategies}

Societies should conduct a research-informed debate on the opportunities and risks of technologies for automated product development and, if necessary, regulate them when their realization seems increasingly imminent. Competition authorities must maintain a functioning, competitive market in which no single player gains too much dominance and control over technologies that may largely determine technological progress. In addition, effects on the labor market should be anticipated. 
When designing automated product development systems and implementing them in an organizational context, work design that benefits employees' needs and work outcomes should be a priority~\supercite{parkerAutomationAlgorithmsWhy2022}.
To reduce the risk of dual-use, access to said systems or their capabilities may be regulated. However, the effectiveness of such measures is questionable and new issues are created.
Strengthening the capabilities of the systems in areas that coincide closely with high-risk applications should not be pursued or only after extensive risk assessment. 
Regulation must ensure that accountability is sufficiently traceable and provable to make laws against illegal use, whether negligent or intentional, enforceable.
Transparency about the use of AI and the associated risks could be increased by requiring an indication of whether products are AI-engineered, or by transferring watermarking techniques, as developed for AI-generated texts~\supercite{kirchenbauerWatermarkLargeLanguage2023} and images~\supercite{wenTreeRingsWatermarksInvisible2023}, to engineering artifacts. 
Furthermore, the lower barrier to innovation is likely to require adjustments to patent laws. Engineering education programs should address the new technologies in their curricula and teach students about their limitations. At the same time, research should inform the development and use of these systems and make claims by system providers verifiable. 
Best practices and regulation needs will evolve alongside the adoption and capabilities of the AI systems. Users, providers and regulatory authorities must understand these developments and adapt their actions accordingly. In both engineering education and business practice, organizational cultures should encourage adherence to good practices. While an ongoing debate on the risks and regulatory needs of such technologies is recommended, some basic principles for safe AI-driven product development can be identified with relative certainty already at this stage.

\subsectionheader{Principles for safe AI-driven product development}

\begin{figure}[t]
    \renewcommand{\figurename}{Box}
    \setcounter{figure}{0}
    {        
        \fontfamily{lmss}\selectfont\small
        \begin{tcolorbox}[title=Principles, title filled=false, colback=mint!10, colframe=blue]
            \vspace{2mm}
            \begin{enumerate}[leftmargin=0.5cm, label=\textbf{\arabic*.}] 
                \item \textbf{Human control and accountability}    
                \begin{enumerate}[label=H\arabic*.]    
                    \item {\textbf{Human control:}} The outer process must be controlled by humans. Otherwise, it would be out of our control.     
                    \item {\textbf{Human accountability:}} Humans are responsible for supervision and are accountable. Otherwise, they are more likely to not fulfil their control obligations for economic or other reasons.
                \end{enumerate}
                \item \textbf{Verifiable design results}
                \begin{enumerate}[label=V\arabic*.]  
                    \item {\textbf{Explainable design:}} The results of the design system must be sufficiently understandable for the responsible actors to enable them to assess safety. Otherwise, they cannot fulfill their supervisory responsibility. 
                    \item {\textbf{Tested design:}} Because theoretical understanding is error-prone and rarely complete, thorough testing is needed to further characterize the design results and verify or extend the responsible actors' understanding.
                \end{enumerate}
                \item \textbf{Strictly confined operating and solution space}
                \begin{enumerate}[label=C\arabic*.]
                    \item {\textbf{Constrained design:}} To minimize the number of potential errors and the attack surface in the event of system manipulation, the task, its requirements and objectives should be narrowly defined and their compliance checked. 
                    \item {\textbf{Sandboxed systems:}} To avoid risks of excessive agency, the design stages and systems operating within them should be separated and information flow restricted.
                \end{enumerate}
                \item \textbf{Systems attuned to humans and the holistic context}
                \begin{enumerate}[label=A\arabic*.]
                    \item {\textbf{(Iterative) Alignment:}} Systems should be well aligned with human intentions, values, and judgments and ask for clarification when in doubt.
                    \item {\textbf{Holistic design:}} The design should be developed and evaluated considering its holistic context.
                \end{enumerate}
                \item {\textbf{Continuous discussion and evolution of community norms and best practices}}
              \end{enumerate}
              \vspace{-1.5mm}
        \end{tcolorbox}
    }
    \caption{Basic principles for safe AI-driven product development}
    \label{box:principles}
\end{figure}

New technologies can change the way products are developed and hence necessitate new best practices and guidelines. For example, in response to the advent of additive manufacturing, guidelines for the ``Design for Additive Manufacturing~(DfAM)'' have been developed~\supercite{thompsonDesignAdditiveManufacturing2016a}. Similarly, if capable AI systems for product development are introduced, it is likely that best practices will emerge in response, providing guidance on how to leverage the new opportunities whilst minimizing the risks. In recent years, several conceptual frameworks for AI-assisted engineering design have been proposed~\supercite{williamsDesignArtificialIntelligence2022,gopfertOpportunitiesLargeLanguage2024}, however, not from a perspective of risk mitigation. With this perspective paper, we aim to start the discussion and consensus-building process on best practices for risk-aware AI-driven product development. 
While regulatory frameworks stipulate product safety, for example, by requiring internal risk analyses and processes to ensure compliance~\supercite{euRegulationEU20232023}, they do not address what constitutes good practices during product development or implications that extend beyond pure product safety (e.g., systematic biases during development), and only refer to unsafe products, not otherwise poorly designed ones. Similarly, AI regulations and safety research provide legal frameworks, best practices, and guidelines for developing and applying AI systems, but they do not take the perspective of engineering design and product development.
Therefore, we propose a set of basic principles for the design of safe systems for automated product development and their safe use (see Box~\ref{box:principles}). They complement and partly inherit best practices of traditional product safety and information security as well as domain-agnostic measures to mitigate the risks of AI systems, such as AI governance~\supercite{tallbergGlobalGovernanceArtificial2023,euRegulationEU20242024}, external auditing~\supercite{brundageTrustworthyAIDevelopment2020,falcoGoverningAISafety2021,mokanderAuditingLargeLanguage2024}, monitoring~\supercite{whittlestoneWhyHowGovernments2021,naihinTestingLanguageModel2023}, red teaming~\supercite{brundageTrustworthyAIDevelopment2020,ganguliRedTeamingLanguage2022}, adversarial testing~\supercite{goodfellowMakingMachineLearning2018,gilmerMotivatingRulesGame2018}, sociotechnical safety evaluation~\supercite{weidingerSociotechnicalSafetyEvaluation2023}, alignment~\supercite{christianoDeepReinforcementLearning2017}, greater transparency (e.g., model cards~\supercite{mitchellModelCardsModel2019}, dataset cards~\supercite{gebruDatasheetsDatasets2021}, transparent post-training~\supercite{casperOpenProblemsFundamental2023}, and incidents disclosure~\supercite{brundageTrustworthyAIDevelopment2020}), watermarking~\supercite{kirchenbauerWatermarkLargeLanguage2023,wenTreeRingsWatermarksInvisible2023}, differential privacy~\supercite{abadiDeepLearningDifferential2016}, and incentives such as `bias or safety bounties'~\supercite{brundageTrustworthyAIDevelopment2020}. 

\begin{figure}[t]
    \centering    \frame{\includegraphics[width=0.89\linewidth]{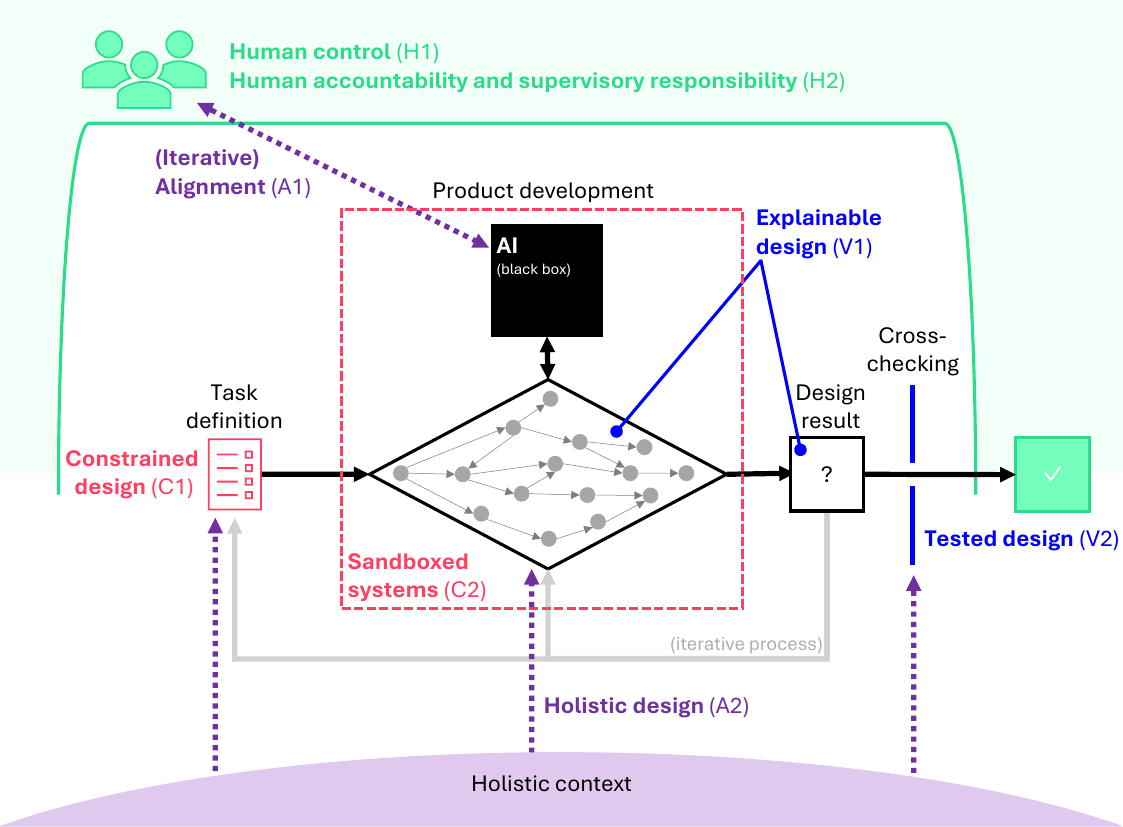}}
    \caption{Principles for safer AI-driven product development (see Box~\ref{box:principles}) situated in an abstract development process.}
    \label{fig:product_development_process_and_rules_for_AI_driven_development}
\end{figure}

Below, we provide additional reasoning to support the principles outlined in Box~\ref{box:principles}. They are visualized in an abstract product development process in Figure~\ref{fig:product_development_process_and_rules_for_AI_driven_development} and mapped to the risks they mitigate alongside the other proposed mitigation strategies in Table~\ref{tab:technical_risks_mitigation_strategy_matrix}.

\paragraph{Human control and accountability} 
The need for human oversight is widely acknowledged~\supercite{aihlegEthicsGuidelinesTrustworthy2019,euRegulationEU20242024}. When discussing the safety of systems that transfer decision making from humans to computers and perform autonomous actions, it is clear that humans must remain in control of these systems and the scope of their application. Thus, the overall design process must be controlled by humans~(H1). Otherwise, it would be out of our control. 
Furthermore, humans are responsible for supervision and are accountable~(H2). Otherwise, they are more likely to not fulfil their control obligations for economic or other reasons.

\paragraph{Verifiable design results}
How can we trust products and technologies that are placed in our hands by an intransparent AI system? Well, if we understand the product or technology well enough, we can trust our understanding of it instead of the black box contriving it. Therefore, explainable design~(V1) should be a central principle of AI-driven product development. The results of design systems must be understandable to the responsible actors to an extent that enables them to assess the safety of the results and fulfil their supervisory responsibilities.
If a design result eludes this degree of understanding, its safety and potential consequences cannot be sufficiently determined. Explainable AI aims to make AI algorithms more understandable. Analogously, if not only the AI algorithms constitute black boxes, but also the products they design become black boxes, explainable design aims to make the products transparent and understandable. Although not a requirement for assessing the safety and quality of a design result, a transparent design process facilitates oversight and improves the interpretability of results.

While sufficient understanding is necessary, human thinking is fallible. Therefore, the design results must be tested~(V2) to characterize it and to inform and verify the responsible actors' understanding. Relying solely on theoretical considerations or empirical evidence leads to errors and incomplete characterizations. Both are required. Should AI-driven product development lessen our understanding of technologies in the long run, empirical testing will become increasingly important to compensate. Although humans must be involved in criticizing and evaluating design results, they may be supported by critique models and digital test environments.

\paragraph{Strictly confined operating and solution space}
A narrowly constrained solution space and clearly defined objectives~(C1) reduce the number of potential errors as well as the attack surface in the event of manipulation. Additionally, it enables more targeted validation. With greater reliance on AI, a more comprehensive specification of constraints and objectives becomes more important, as what is obvious to humans is not necessarily obvious to AI. Although striving for optimality is somewhat intrinsic to engineering design, and design automation systems will include some sort of optimization process, an increased focus on optimality is valuable in evaluating automatically generated designs because it reduces the space of acceptable solutions while inherently adding value. In other words, if a design solution is knowingly non-optimal or unintuitive with respect to the defined constraints and objectives, it should be given careful consideration. 
For example, a design should not comprise unnecessary features that do not contribute to the design objective, as they may have a harmful secondary function. Of course, also technologies that are perfectly designed for their intended use are prone to dual-use. Furthermore, the pursuit of optimality can have negative effects if false constraints or misleading proxy objectives are set.
 
An invalid solution can be made valid by altering the design objectives, constraints or validation.
Therefore, design task setting, its execution, and validation must be separated. It follows that the design generation system must not be authorized to cross-check its own designs or alter the task by changing requirements, objectives, and so forth. 
The permissions and allowed information flows of the design system and other agentic systems used in the design process must be carefully considered~(C2).
A balance must be struck between restricting the flow of information and allowing an iterative design process. Furthermore, it follows that generated designs are not allowed to convey hidden messages from the design model to other models. For example, the systems must prevent prompt injections using component naming or engravings in CAD models, among other things.
\renewcommand{\bottomfraction}{.96}
\begin{table}[bth]
    \centering
    \hspace{-1.2cm}
    \includegraphics[width=0.94\linewidth]{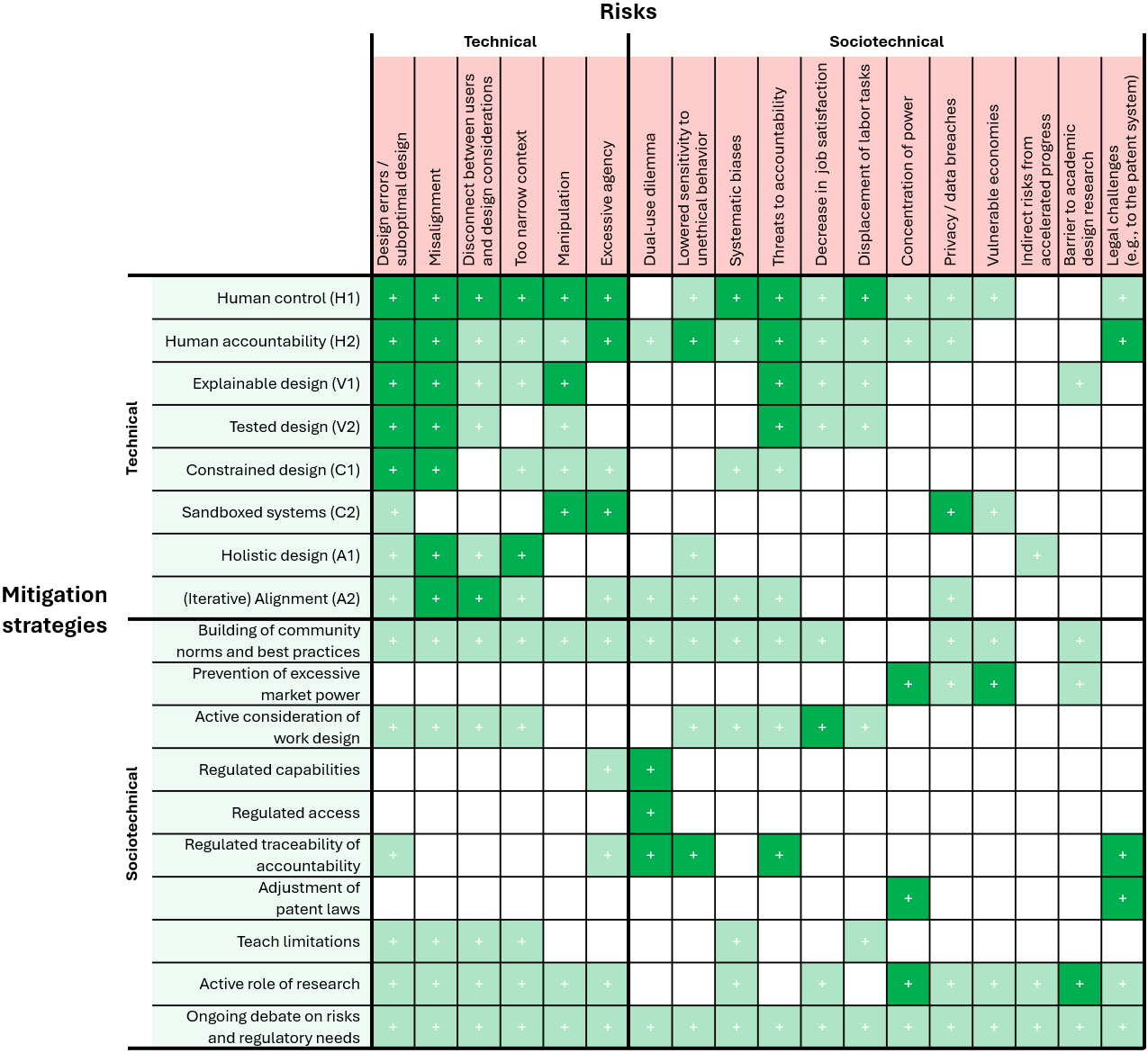}
    \caption{Relationship between risks and mitigation strategies. The technical risks and mitigation strategies are mostly related to design quality and the safe use of automated product development systems at the individual level. The sociotechnical perspective goes beyond this and considers risks that aggregate at the societal level. A green cell indicates that the strategy contributes strongly (dark) or moderately (light) to mitigating the corresponding risk. As the mapping is based on the judgment of the authors, it should be taken with caution.}  \label{tab:technical_risks_mitigation_strategy_matrix}
\end{table}

\paragraph{Systems attuned to humans and the holistic context}
Good design takes human values and preferences into account. At the same time, it is important that autonomous systems correctly interpret user intentions. Therefore, such systems need to be well aligned with human intentions, values, and judgments~(A1). Arguably, this is helped by asking for clarification when in doubt. The fact that even close people occasionally misunderstand each other illustrates that ``grounding'' of AI can never be solved absolutely, and that ``alignment'' must hence be approached dynamically by asking for clarification.

Although a narrow solution space is sought, the design should be developed and evaluated considering its holistic context~(A2). This is not a contradiction. Human engineers design products in context of their world knowledge and using common sense. Together they already provide many guardrails for the design space. However, as human world knowledge and common sense are both very profound and universal, it is difficult to attest an AI similar world knowledge and sense making. Therefore, it is the engineers responsibility to assess the design in a more holistic context and system providers are urged to address this issue. For example, when designing a rain jacket in an isolated technical context, one might be tempted to use toxic ``forever chemicals'' in the coating to improve its performance, but one would refrain from doing so when considering human health and the environment. 
A reasonable definition of constraints limits the risks of neglecting the holistic context, but cannot replace it, as constraints tend to be incomplete. 

The proposed principles are only a basis and should be continually discussed, adapted, and expanded as technology, its application, and our understanding evolve.

\sectionheader{Conclusion}
With this perspective paper, we aim to initiate a discussion and consensus-building process regarding the risks associated with AI-driven product development and how to mitigate them.

By potentially accelerating innovation by increasing engineering capabilities and/or lowering development cost whilst re-distributing inventive power and labor, the introduction of systems that automate large parts of product development will likely have a major impact on our socioeconomic system. 
The consolidation towards a few dominant players of such systems would pose an immense risk, as a monopoly on inventions is qualitatively different from a monopoly on some consumer good.
In addition, the potential shift of power from workers to companies has many implications. If companies rely more on technology and less on people, this will affect the labor market, job satisfaction, introduce widespread systematic biases, and make them less sensitive to unethical developments. An economy's dependence on AI systems for research and development makes it vulnerable and the systems a target for criminals and foreign governments alike. Furthermore, the ``democratization'' of product development entails the risk of the dual-use dilemma. Beyond that there may be indirect risks, such as environmental damage from increased consumption.

At the same time, the introduction of such systems will affect design quality and safety. Reasons for this can be insufficient capabilities of the systems combined with insufficient human oversight and judgment, misalignment between the design objectives as interpreted by the AI systems and the user's intent, their operation in a too narrow context, their vulnerability to manipulation and excessive agency, and the detachment of product development from the people it should benefit.

To mitigate these risks societies need to conduct a research-informed debate on the opportunities and risks of technologies for automated product development and, if necessary, regulate them when their realization seems increasingly imminent. Competition authorities must prevent market dominance. A transforming job market should be anticipated and patent law may need adaptation. 
Engineers should be educated about best practices and the systems' limitations. Work design that benefits employees' needs and work outcomes should be considered in the development and implementation of these systems. While an ongoing debate on the risks and regulatory needs of such technologies is recommended, some basic principles for the development of safer systems for automated product development and their safer use can be identified with relative certainty already at this stage, namely human control~(H1), human accountability and supervisory responsibility~(H2), explainable design~(V1), tested design~(V2), narrowly constrained design~(C1), sandboxing~(C2), (iterative) alignment~(A1), and holistic design~(A2).

Further research should involve a wider group of stakeholders and use systematic methods to reach consensus. As the considerations in this perspective are theoretical, empirical research is needed to substantiate or refute them. Besides further research on the risks and regulatory needs, there are many interesting research directions. This paper does not cover the opportunities associated with using AI in product development and the wide array of neutral and positive consequences. For example, the ``not invented here'' tendency may decrease if product development becomes less associated with human ingenuity. Future research should weigh the risks and opportunities.
Furthermore, research is needed to provide practical guidance to organizations that are implementing AI in their development teams today. How engineers will use this technology has a major impact on its assessment and remains to be explored. In this perspective, we assume agent-based systems that rely on foundation models similar to today’s LLMs while future systems may differ. We advocate the development of transparent systems that keep engineers in control and facilitate their work, rather than fixating on a grand vision of autonomous R\&D capabilities.

Although technology for AI-driven product development is still in its infancy, we hope that starting a debate about its opportunities and risks now will contribute to balancing them without dangerously long delays in understanding, building of community norms, and regulation.

\bigskip
\sectionheader{References}
\printbibliography[heading=none]

\section*{Acknowledgements}

The authors would like to thank the German Federal Government, the German State Governments, and the Joint Science Conference~(GWK) for their funding and support as part of the NFDI4Ing consortium. Funded by the German Research Foundation~(DFG) -- project number: 442146713. Furthermore, this work was supported by the Helmholtz Association under the program “Energy System Design”.

\section*{Author contributions statement}
J.G. wrote the original draft. J.W., P.K., and N.P. revised the draft. P.K., J.W., N.P., and J.L. took care of the supervision, project administration, and funding acquisition. All authors reviewed the manuscript.

\section*{Competing interests}

The authors declare no competing interests.

\end{document}